\documentclass[12pt,a4paper]{article}
\usepackage{latexsym}
\usepackage{bbm}
\usepackage{amssymb}
\usepackage{amsmath}

\renewcommand{\title}[1]{\null\vspace{25mm}\noindent{\Large{\bf
#1}}\vspace{10mm}}
\newcommand{\authors}[1]{\noindent{\large #1}\vspace{20mm}}
\newcommand{\address}[1]{{\center{\noindent #1\vspace{0mm}}}}
\renewcommand{\abstract}[1]{\vspace{17mm}
\noindent{\small{\em Abstract.} #1}\vspace{2mm}}

\begin{document}                

\begin{titlepage}

\begin{center}
\hspace*{\fill}{{\normalsize \begin{tabular}{l}
                              \textsf{hep-th/0202092}\\
                              \textsf{REF. TUW-02-02}\\
                             \end{tabular}   }}

\title{The role of the field redefinition in noncommutative Maxwell theory}

\vspace{10mm}

\authors{  \large{I. Fr\"{u}hwirth\footnote{fruhwirth@hep.itp.tuwien.ac.at},
         J. M. Grimstrup\footnote{Work supported by The Danish Research Agency. jesper@hep.itp.tuwien.ac.at},
  Z. Morsli\footnote{morsli@hep.itp.tuwien.ac.at},\\
 L. Popp\footnote{Work supported in part by ``Fonds zur F\"orderung der Wissenschaftlichen Forschung'' (FWF) under contract P13125-PHY. popp@hep.itp.tuwien.ac.at},
  M. Schweda\footnote{mschweda@tph.tuwien.ac.at}
                 }     
        }

\vspace{5mm}

\address{Institut f\"ur Theoretische Physik, Technische
Universit\"at Wien\\
      Wiedner Hauptstra\ss e 8--10, A-1040 Wien, Austria}

\end{center} 
\thispagestyle{empty}
\begin{center}
\begin{minipage}{12cm}

\vspace{10mm}

{\it Abstract.} We discuss
$\theta$-deformed 
Maxwell theory at first order in $\theta$ with the help of the Seiberg-Witten (SW) map. With an appropriate field redefinition consistent with the
SW-map we analyse the one-loop corrections of the vacuum polarization of
photons. We show that the radiative corrections obtained in a previous work may
be described by the Ward-identity of the BRST-shift symmetry corresponding  
to a field redefinition.

\setcounter{footnote}{0}

\end{minipage}\end{center}
\end{titlepage}

\section{Introduction}

Discussing noncommutative quantum field theories, especially noncommutative
gauge field models, one has in principle many possibilities to formulate such
models. The traditional approach widely used in many papers, 
\cite{Moyal:sk}, \cite{Hayakawa:1999zf}, \cite{Hayakawa:1999yt},
\cite{Matusis:2000jf}, is based on the fact that  noncommutative gauge
theory is realized as a theory on the set of ordinary 
functions by modifying the product of two functions in terms of the $\star$-product \cite{Moyal:sk}. 
The simplest candidate would be a noncommutative counterpart of QED. Due to the
noncommutativity
also the noncommutative $U(1)$ gauge field model (with and without matter)
has a non-Abelian structure implying
the usual BRST-quantization procedure involving the appearence of
$\phi\pi$-ghost fields.

Unfortunately, the perturbative realization of such gauge field models develops
unexpected new features: the so called UV/IR mixing emerging from nonplanar
graphs. Explicitly, this can be seen by computing one-loop corrections of the
vacuum polarization of the photon \cite{Hayakawa:1999zf}, \cite{Hayakawa:1999yt}. 
The evaluation of the finite part of the 
gauge field vacuum polarization shows the
existence of singularities in the infrared limit. Those IR-singularities forbid
Feynman graph computation of higher loop orders. Presently, one believes that
such IR-singularities can perhaps be avoided by an appropriate field
redefinition \cite{x}.

A second formulation of noncommutative gauge field models 
simultaneously incorporates the deformed product and 
the Seiberg-Witten (SW) map.
 The SW-map ensures the gauge equivalence between an ordinary
gauge field model and its noncommutative counterpart 
\cite{Seiberg:1999vs}, \cite{Madore:2000en}, \cite{Jurco:2000ja},
\cite{Jurco:2001rq}. In 
this sense it is possible to expand the noncommutative gauge field as series in
the ordinary gauge field and the deformation parameter of the noncommutative
space-time geometry $\theta^{\mu\nu}$.

Additionally, one  has  the possibility to formulate gauge
field models on noncommutative spaces via covariant coordinates 
\cite{Madore:2000en}, \cite{Jurco:2000ja}, \cite{Jurco:2001rq}. Also in these
approaches the use of the 
SW-map emerges quite naturally.

The present paper is devoted to study 
$U(1)$ noncommuta\-tive Yang-Mills ($U(1)$-NCYM) theory in the context of the SW-map
for the simplest case allowing only a linear dependence on the deformation
parameter $\theta^{\mu\nu}$. The corresponding
$U(1)$-deformed gauge invariant action has been derived in \cite{Jurco:2001rq},
\cite{Bichl:2001nf} and  recently  been used very often to study physical
consequences  of such a deformed
Maxwell theory \cite{Kruglov:2001dm}, \cite{Jackiw:2001dj},
\cite{Guralnik:2001ax}, \cite{Cai:2001az}.

It is remarkable to note that in this simple $\theta$-deformed Maxwell theory
(in its perturbative realization) at the one-loop level no IR-singularities in
the above sense exist \cite{Bichl:2001nf}, \cite{Bichl:2001gu}. As
 is further explained in \cite{Bichl:2001cq} a consistent field redefinition
(in agreement with the SW-map) allows to add to the gauge field of the
$\theta$-deformed Maxwell theory further $\theta$-dependent and gauge invariant
terms. Such additional terms are very useful for studying one-loop corrections
of the vacuum polarization of the photon \cite{Bichl:2001cq}. 
Since the interaction contains terms linear 
in $\theta$ and trilinear in the Abelian field strength one obtains terms
proportional to the square of $\theta$ in the corresponding vacuum polarization
at the one loop level.
However, by redefining the relevant unphysical free
parameters of the above mentioned field redefinition one is able to carry out
the renormalization procedure in the usual sense at least for this simple case.
More recently, it has been argued that the linear 
$\theta$-deformed QED (with the inclusion of fermions) leads to a
nonrenormalizable theory \cite{Wulkenhaar:2001sq} at the one-loop level.

The aim of this paper is to focus on the connection between
the above mentioned field redefinition and the so-called BRST-shift symmetry
described in \cite{Bastianelli:1990ey}, \cite{Alfaro:1992cs}. Especially, we try to explain that
the perturbative corrections of the photon self-energy are compatible with the
Ward-identity (WI) of the shift symmetry.

The paper is organized as follows: Sec.~\ref{2} is concerned with the presentation of
the simplest $\theta$-deformed Maxwell theory (in terms of the usual Abelian
gauge symmetry of the photon field). This is realized by starting with 
$U(1)$-NCYM and using the SW-map to lowest order
in $\theta^{\mu\nu}$. In a next step one performs a field redefinition in
consistency with the SW-map in order to derive the physical consequences leading
to the WI for the BRST-shift symmetry.

In order to understand this BRST-shift symmetry we assume the existence of an
Abelian gauge invariant action with an appropiate gauge fixing implemented by a
 multiplier field B:
\begin{equation}\label{def-1}
\Gamma^{(0)}[A,B]=\Gamma_\mathrm{inv}[A]+\Gamma_\mathrm{gf}[B,A].
\end{equation}
The description of the shift symmetry is based on the existence of the
following redefinition
\begin{equation}\label{def-2}
A_{\mu }=\tilde{A}_{\mu }+ \mathbbm{A} ^{(2)}_{\mu }(\tilde{A}),
\end{equation}
where the upper index indicates quadratic dependence on $\theta^{\mu\nu}$
which will be needed in the future.

The derivation of the BRST-shift symmetry needs a new
type of gauge fixing besides $\Gamma_\mathrm{gf}[B,A]$. Here 
we follow \cite{Bastianelli:1990ey}, \cite{Alfaro:1992cs} adapted for our special
case\footnote{
In consistency with the formulation of noncommutative gauge field models we use
here $\star$-products. 
}
 by
defining:
\begin{align}\label{def-3}
\Sigma &= \Gamma ^{0}[A_\mu,B]+\Gamma_\mathrm{shift}+\Gamma_\mathrm{gf-shift}
\\ \nonumber 
&= \Gamma ^{0}[A_\mu,B]+
\int d^{4} x \int d^{4} y \, \bar{c}^{\mu }(x)\star\frac{\delta \varphi_{\mu
}(x)}{\delta \tilde{A}_{\sigma }(y)}\star c_{\sigma }(y) 
\\ \nonumber 
&\phantom{=}+ \int d^{4} x \, \Pi^\mu \star \Big(\varphi_\mu-A_\mu\Big),
\end{align}
where 
\begin{equation}\label{def-4}
\varphi_\mu=\tilde{A}_\mu+\phi_\mu^{(2)}(\tilde{A}).
\end{equation}
The vectorial antighost $\bar{c}^{\mu }$ and ghost field $c^{\mu }$ carry a
Faddeev-Popov ghost charge of $-1$ and $+1$, respectively.

Using (\ref{def-2}) and (\ref{def-4}) one gets
\begin{multline}\label{def-5}
\Sigma = \Gamma ^{0}[A_\mu,B]+ \int d^{4} x \, \bar{c}^{\mu }(x)c_{\mu}(x)
+\int d^{4} x \int d^{4} y \, \bar{c}^{\mu }(x)\frac{\delta  \phi^{(2)} _{\mu
}(x)}{\delta \tilde{A}_{\sigma }(y)}c_{\sigma }(y)
\\ 
+ \int d^{4} x \, \Pi^\mu \Big(\phi^{(2)}_\mu-\mathbbm{A}^{(2)}_\mu(\tilde{A})\Big).
\end{multline}
Since $\phi_\mu^{(2)}(\tilde{A})$ and
$\mathbbm{A}_\mu^{(2)}(\tilde{A})$ are of order $2$ in $\theta$ no $\star$-products are
needed. 
The BRST-shift symmetry is defined as
\begin{equation}\label{def-6}
s\bar{c}^{\mu } = -\Pi_\mu,\quad
s\tilde{A}_\mu = c_\mu = -s\mathbbm{A}^{(2)}_\mu, \quad
sA_\mu = s\Pi_\mu = sc_\mu = 0.
\end{equation}
These transformations are manifestly nilpotent. However,  eliminating the
multiplier field $\Pi_\mu$ via an algebraic equation of motion yields
\begin{equation}\label{def-7}
\frac{\delta  \Sigma}{\delta \Pi_\mu} =
\phi^{(2)}_\mu - \mathbbm{A}^{(2)}_\mu(\tilde{A}) = 0,
\end{equation}
and, additionally, from (\ref{def-3}) follows
\begin{equation}\label{def-8}
\frac{\delta  \Sigma}{\delta A_\mu} =
\frac{\delta  \Gamma^{(0)}}{\delta A_\mu} - \Pi_\mu = 0.
\end{equation}
This implies for the transformations (\ref{def-6})
\begin{equation}\label{def-9}
s\bar{c}^{\mu } = -\;\frac{\delta  \Gamma^{(0)}}{\delta A_\mu},\quad
s\tilde{A}_\mu = c_\mu,\quad
sc_\mu = 0.
\end{equation}
These equations show explicitly that off-shell nilpotency is lost and that
the shift symmetry is no longer linear.

One has to introduce an unquantized external source to describe the nonlinear
transformation of $s\bar{c}^{\mu }$. 
 This will be used for the
construction of the corresponding WI for the shift symmetry in sec.~\ref{2}.

\section{Deformed Maxwell theory---Consequences of the BRST-shift symmetry at
the classical level}\label{2}

In order to derive the corresponding
$\theta$-deformed Maxwell theory one starts with the $U(1)$-NCYM
 model
\cite{Madore:2000en}, \cite{Bichl:2001nf}: 
\begin{equation}\label{def-2-1}
\Gamma _\mathrm{inv}^{(0)}=-\frac{1}{4}\int d^{4} x \, \hat{F}_{\mu \nu }
\star \hat{F}^{\mu \nu},
\end{equation}
where the field strength in terms of the noncommutative $U(1)$ gauge field
$\hat{A}_{\mu}$ is given by
\begin{equation}\label{def-2-2}
\hat{F}_{\mu \nu }=\partial_\mu\hat{A}_{\nu }-\partial_\nu\hat{A}_{\mu }
-i\Big[\hat{A}_{\mu },\hat{A}_{\nu }\Big]_M.
\end{equation}
The Moyal bracket is defined by
\begin{equation}\label{def-2-3}
\Big[\hat{A}_{\mu },\hat{A}_{\nu }\Big]_M=
\hat{A}_{\mu }\star \hat{A}_{\nu }-\hat{A}_{\nu }\star \hat{A}_{\mu },
\end{equation}
 using the $\star$-product 
\begin{equation}\label{def-2-4}
A(x)\star B(x)
=e^{\frac{i}{2}\theta ^{\mu \nu }\partial _{\mu }^{\alpha }\partial _{\nu }^{\beta }}
A(x+\alpha )B(x+\beta )\Big |_{\alpha =\beta =0},
\end{equation}
where $\theta ^{\mu \nu }$ is the deformation parameter of the noncommutative
geometry.

The action (\ref{def-2-1}) is invariant under the infinitesimal noncommutative
gauge transformation
\begin{equation}\label{def-2-5}
\hat{\delta}_{\hat{\lambda}} \hat{A}_\mu = 
\partial_\mu \hat{\lambda} - i \hat{A}_\mu \star \hat{\lambda} 
+ i \hat{\lambda} \star \hat{A}_\mu 
\equiv \hat{D}_\mu \hat{\lambda}.
\end{equation}
It was shown by Seiberg and Witten \cite{Seiberg:1999vs} that an expansion in
$\theta ^{\mu \nu }$ leads to a map between the noncommutative gauge field
$\hat{A}_\mu$ and the commutative gauge field $A_\mu$ as well as their
respective gauge parameters $\hat{\lambda}$ and $\lambda$, known as the SW-map. To
lowest order in $\theta$ one has 
 in  the Abelian case \cite{Seiberg:1999vs}, \cite{Madore:2000en}, \cite{Bichl:2001nf}
\begin{align}\label{def-2-6}
\hat{A}_{\mu }(A_{\mu }) &= A_{\mu }-\frac{1}{2}A_{\rho}
\Big (\partial _{\sigma}A_{\mu }+F_{\sigma \mu }\Big )
+ O(\theta^2),
\\ \nonumber
\hat{\lambda} (A_{\mu },\lambda)
&= \lambda -\frac{1}{2}\theta ^{\rho \sigma}A_{\rho}\partial _{\sigma}\lambda
+ O(\theta^2).
\end{align}
In (\ref{def-2-6}) $F_{\sigma \mu }$ is the ordinary Abelian field strength
given by
\begin{equation}\label{def-2-6a}
F_{\sigma \mu } = \partial _{\sigma}A_{\mu } - \partial _{\mu}A_{\sigma}.
\end{equation}
Using (\ref{def-2-1}), (\ref{def-2-2}), (\ref{def-2-3}), (\ref{def-2-4}) and
(\ref{def-2-6}) one gets to lowest order in $\theta ^{\mu \nu }$
\cite{Jurco:2001rq},  \cite{Bichl:2001nf}
\begin{equation}\label{def-2-7}
\Gamma _\mathrm{inv}=\int d^4 x \, \Big (-\frac{1}{4}F_{\mu \nu }F^{\mu \nu
}-\frac{1}{2}\theta ^{\rho \sigma }\Big (F_{\mu \rho }F_{\nu \sigma }F^{\mu \nu
}-\frac{1}{4}F_{\rho \sigma }F_{\mu \nu }F^{\mu \nu }\Big )\Big )
+ O(\theta^2),
\end{equation}
which is invariant under the usual Abelian gauge transformation
\begin{equation}\label{def-2-8}
\delta A_{\mu } = \partial _{\mu} \lambda.
\end{equation}
The action (\ref{def-2-7}) has in its full form, involving all orders of 
$\theta^{\mu \nu }$, infinitely many interactions of infinitely high order in
the field. Additionally, since $\theta^{\mu \nu }$ has dimension $-2$, the model
is power-counting nonrenormalizable in the traditional sense.

In order to quantize the model one introduces a Landau gauge fixing
\begin{equation}\label{def-2-9}
\Gamma _\mathrm{gf}=\int d^4 x \, B\partial ^{\mu }A_{\mu },
\end{equation}
where $B$ is the multiplier field implementing the gauge $\partial ^{\mu }A_{\mu
}=0$.

Then one establishes the shift symmetry according to
\cite{Bastianelli:1990ey}, \cite{Alfaro:1992cs}. As  is explained in \cite{Bichl:2001cq} the
relevant  field redefinition, compatible with the SW map, takes the
following form:
\begin{equation}\label{def-2-10}
A_{\mu }=\tilde{A}_{\mu }+ \mathbbm{A} ^{(2)}_{\mu }(\tilde{A}),
\end{equation}
where the upper index again indicates that \(  \mathbbm{A} ^{(2)}_{\mu }(\tilde{A}) \)
depends quadraticly on \( \theta^{\mu\nu}  \).
Additionally, \(  \mathbbm{A} ^{(2)}_{\mu }(\tilde{A}) \) is gauge invariant with
respect to (\ref{def-2-8})
\begin{equation}\label{def-2-11}
\delta_\lambda  \mathbbm{A} ^{(2)}_{\mu }(\tilde{A}) = 0.
\end{equation}
Terms linear in \( \theta^{\mu\nu}  \) are
excluded due to the topological nature of the corresponding action.\footnote{In \cite{Bichl:2001cq} one finds the most general 
$\mathbbm{A} ^{(2)}_{\mu }(\tilde{A})$.}

On the other hand the formula (\ref{def-2-10}) allows the introduction of
terms with a quadratic dependence  already in the classical action. Such terms
are needed 
for the one-loop renormalization procedure of the vacuum polarization of
photons.

In the spirit of \cite{Bastianelli:1990ey}, \cite{Alfaro:1992cs} one defines now
for the 
deformed Maxwell theory, eq.~(\ref{def-2-7}), the corresponding shift-action in
the following way, see  formula (\ref{def-5})
\begin{align}\label{def-2-12}
\Gamma _\mathrm{shift}&=
\int d^{4}x\int d^4 y \, \bar{c}^{\mu }(x)\star\frac{\delta A_{\mu }(x)}{\delta
\tilde{A}_{\sigma }(y)}\star c_{\sigma }(y) 
\\ \nonumber
&=\int d^4 x \, \bar{c}^{\mu }(x)c_{\mu}(x)
+\int d^{4}x\int d^4 y \, \bar{c}^{\mu }(x)\frac{\delta  \mathbbm{A}^{(2)} _{\mu
}(x)}{\delta \tilde{A}_{\sigma }(y)}c_{\sigma }(y), &  & 
\end{align}
where $\bar{c}^{\mu }$ and $c^{\mu }$  are the vectorial shift ghost and
antighost fields, respectively. Due to the
fact that $\mathbbm{A}^{(2)} _{\mu}$ is already of second order in $\theta$ one
can neglect the stars in the second term of (\ref{def-2-12}).\footnote{In (\ref{def-2-12}) we have used 
$\int d^4 x \, \bar{c}^{\mu}(x) \star c_{\mu}(x) 
 = \int d^4 x \, \bar{c}^{\mu }(x)c_{\mu}(x)$.}

Thus, the total action of the model under consideration ready for quantization
is given by 
\begin{multline}\label{def-2-12a}
\Gamma ^{(0)}=\Gamma ^{(0)} _\mathrm{inv}+\Gamma _\mathrm{gf}+\Gamma _\mathrm{shift} 
\\ 
=\int d^4 x \, \Big (-\frac{1}{4}F_{\mu \nu }F^{\mu \nu }-\frac{1}{2}\theta ^{\rho
\sigma }\Big (F_{\mu \rho }F_{\nu \sigma }F^{\mu \nu }-\frac{1}{4}F_{\rho \sigma
}F_{\mu \nu }F^{\mu \nu }\Big )\Big ) \\  
+\int d^4 x \, B\partial ^{\mu }A_{\mu }
+\int d^4 x \, \bar{c}^{\mu }(x)c_{\mu }(x)+\int d^{4}x\int d^4 y \, \bar{c}^{\mu
}(x)\frac{\delta  \mathbbm{A}^{(2)} _{\mu }(x)}{\delta \tilde{A}_{\sigma
}(y)}c_{\sigma }(y). 
\end{multline}
More explicitly, one has 
\begin{multline}
\label{def-2-13}
\Gamma ^{(0)}=\int d^4 x \, \Big (-\frac{1}{4}\Big (\tilde{F}_{\mu \nu
}\tilde{F}^{\mu \nu } 
-4\partial ^{\mu }\tilde{F}_{\mu \nu } \mathbbm{A} ^{(2)\nu }\Big )
 \\ 
-\frac{1}{2}\theta ^{\rho \sigma }\Big (\tilde{F}_{\mu \rho }\tilde{F}_{\nu
\sigma }\tilde{F}^{\mu \nu }-\frac{1}{4}\tilde{F}_{\rho \sigma }\tilde{F}_{\mu
\nu }\tilde{F}^{\mu \nu }\Big )\Big ) 
+\int d^4 x \, \Big (B\partial ^{\mu }\tilde{A}_{\mu }+B\partial ^{\mu }
\mathbbm{A}^{(2)}_{\mu }\Big ) 
\\ 
+\int d^4 x \, \bar{c}^{\mu }(x)c_{\mu }(x)+\int d^{4}x\int d^4 y \, \bar{c}^{\mu
}(x)\frac{\delta  \mathbbm{A}^{(2)} _{\mu }(x)}{\delta \tilde{A}_{\sigma
}(y)}c_{\sigma }(y), 
\end{multline}
where $\tilde{F}_{\alpha\beta}$ is the usual Abelian field strength in terms of
$\tilde{A}_{\beta}$.

From eq.~(\ref{def-9}) the
BRST-shift symmetry of the action (\ref{def-2-13}) is given by
\begin{align}\label{def-2-14}
s\bar{c}^{\tau }&=\frac{\delta \Gamma ^{(0)}}{\delta A_{\tau }}
\Bigg |_{A_{\mu }=\tilde{A}_{\mu }+ \mathbbm{A} ^{(2)}_{\mu }(\tilde{A})}
 \\ \nonumber
&=\partial _{\mu }F^{\mu \tau }+ 
+\theta ^{\tau \sigma }\partial _{\mu }(F_{\nu \sigma }F^{\mu \nu })-\theta
^{\rho \sigma }\partial _{\rho }(F_{\nu \sigma }F^{\tau \nu })+\theta ^{\rho
\sigma }\partial _{\mu }(F^{\mu }_{\rho }F_{\sigma }^{\tau })
\\ \nonumber 
&\phantom{=}-\frac{1}{4} \theta ^{\rho \tau }\partial _{\rho }\Big (F_{\mu \nu }F^{\mu \nu
}\Big )-\frac{1}{2} \theta ^{\rho \sigma }\partial _{\mu }\Big (F_{\rho \sigma
}F^{\mu \tau }\Big )-\partial ^{\tau }B 
\\ \nonumber
&=\partial _{\mu }(\tilde{F}^{\mu \tau }+ \partial^\mu \mathbbm{A} ^{(2)\tau}
        - \partial^\tau \mathbbm{A} ^{(2)\mu})
+\theta ^{\tau \sigma }\partial _{\mu }(\tilde{F}_{\nu \sigma }\tilde{F}^{\mu
\nu })-\theta ^{\rho \sigma }\partial _{\rho }(\tilde{F}_{\nu \sigma
}\tilde{F}^{\tau \nu })
\\ \nonumber
&\phantom{=}+\theta ^{\rho \sigma }\partial _{\mu }(\tilde{F}^{\mu
}_{\rho }\tilde{F}_{\sigma }^{\tau })
-\frac{1}{4} \theta ^{\rho \tau }\partial _{\rho }\Big (\tilde{F}_{\mu \nu
}\tilde{F}^{\mu \nu }\Big )-\frac{1}{2} \theta ^{\rho \sigma }\partial _{\mu
}\Big (\tilde{F}_{\rho \sigma }\tilde{F}^{\mu \tau }\Big )-\partial ^{\tau }B \\
\nonumber 
&=:\partial _{\mu }(\tilde{F}^{\mu \tau }+ \partial^\mu \mathbbm{A} ^{(2)\tau}
        - \partial^\tau \mathbbm{A} ^{(2)\mu})
        -\partial ^{\tau }B+\mathcal{F}^{(1)\tau }(\tilde{A}), 
\\ \nonumber
s\tilde{A}_\mu &= c_\mu, \\ \nonumber
sc_{\mu} &= sB = 0,
\end{align}
where 
\begin{align}\label{def-2-14a}
\mathcal{F}^{(1)\tau }(\tilde{A}):=
\theta ^{\tau \sigma }\partial _{\mu }(\tilde{F}_{\nu \sigma }\tilde{F}^{\mu \nu
})-\theta ^{\rho \sigma }\partial _{\rho }(\tilde{F}_{\nu \sigma
}\tilde{F}^{\tau \nu })+\theta ^{\rho \sigma }\partial _{\mu }(\tilde{F}^{\mu
}_{\rho }\tilde{F}_{\sigma }^{\tau })
 \\ \nonumber \phantom{:=}
-\frac{1}{4} \theta ^{\rho \tau }\partial _{\rho }\Big (\tilde{F}_{\mu \nu
}\tilde{F}^{\mu \nu }\Big )-\frac{1}{2} \theta ^{\rho \sigma }\partial
_{\mu }\Big (\tilde{F}_{\rho \sigma }\tilde{F}^{\mu \tau }\Big ).
\end{align}
One has to comment at this point that the BRST-shift symmetry for the vectorial
antighost field $\bar{c}^{\mu }$ is highly nonlinear. Additionally, as is explained in the
introduction, the off-shell nilpotency is also lost:
\begin{equation}\label{def-2-15}
s^2\bar{c}^{\mu } \neq 0.
\end{equation}
Since the transformation of the antighost field $\bar{c}^{\mu }$ contains
nonlinear expressions one must introduce an external unquantized source
$\rho_\mu$ for the term $\mathcal{F}^{(1)\mu}(\tilde{A})$. This implies a further piece in the action (\ref{def-2-11})
\begin{equation}\label{def-2-16}
\Gamma_{ext}=\int d^4 x \,  \rho_\mu \mathcal{F}^{(1)\mu}(\tilde{A}),
\end{equation}
where $\rho_\mu$ is gauge invariant.

The new total action becomes therefore
\begin{align}\label{def-2-17}
\Gamma ^{(0)}&=\int d^4 x \, \Big (-\frac{1}{4}\Big (\tilde{F}_{\mu \nu }
        \tilde{F}^{\mu \nu }
        -4\partial ^{\mu }\tilde{F}_{\mu \nu } \mathbbm{A} ^{(2)\nu }\Big)
\\ \nonumber    
        &\phantom{=}-\frac{1}{2}\theta ^{\rho \sigma }\Big (\tilde{F}_{\mu \rho }
        \tilde{F}_{\nu\sigma }\tilde{F}^{\mu \nu }
        -\frac{1}{4}\tilde{F}_{\rho \sigma }\tilde{F}_{\mu\nu }
        \tilde{F}^{\mu \nu }\Big )\Big ) 
+\int d^4 x \, \Big (B\partial ^{\mu }\tilde{A}_{\mu }+B\partial ^{\mu }
\mathbbm{A}^{(2)} _{\mu }\Big ) \\ \nonumber 
        &\phantom{=}+\int d^4 x \, \bar{c}^{\mu }(x)c_{\sigma }(x)+\int d^{4}x\int d^4 y \, \bar{c}^{\mu
}(x)\frac{\delta  \mathbbm{A}^{(2)} _{\mu }(x)}{\delta \tilde{A}_{\sigma }(y)}c_{\sigma }(y)
\\ \nonumber
        &\phantom{=}+\int d^4 x \,  \rho_\mu \mathcal{F}^{(1)\mu}(\tilde{A}). 
\end{align}
Now we are able to characterize the symmetry content of the BRST-shift symmetry
by the following nonlinear WI:
\begin{multline}\label{def-2-18}
\mathcal{S}(\hat{\Gamma} ^{(0)}) = \\
\int d^{4} x \, \Big (\Big (\partial _{\rho }(\tilde{F}^{\rho \mu}
        + \partial^{\rho} \mathbbm{A}^{(2)\mu}
        - \partial^\mu \mathbbm{A} ^{(2)\rho}) 
 - \partial^\mu B+\frac{\delta \hat{\Gamma}^{(0)}}{\delta \rho^\mu} \Big)
 \frac{\delta\hat{\Gamma}^{(0)}}{\delta \bar{c}^\mu}
  +c_\mu \frac{\delta\hat{\Gamma}^{(0)}}{\delta \tilde{A}^\mu}\Big)
= 0.
\end{multline}
Eq.~(\ref{def-2-18}) will be the key for the understanding of the
radiative corrections of the 2-point vertex-functional at the one-loop-level 
\cite{Bichl:2001cq}.

Additionally, our model is also characterized by the gauge symmetry
(\ref{def-2-8}) with
\begin{equation}\label{def-2-19}
\delta_\lambda \bar{c}^\mu = \delta_\lambda c^\mu = \delta_\lambda B = 0.
\end{equation}
This ordinary gauge invariance is described by the following
WI operator
\begin{equation}\label{def-2-20}
W_\lambda=\int d^4 x \,  \partial_\mu \lambda 
\frac{\delta}{\delta A_\mu(x)}
\end{equation}
and, as usual, the gauge symmetry is broken by the gauge fixing. This leads to
\begin{equation}\label{def-2-21}
W_\lambda \hat{\Gamma}^{(0)}
=\int d^4 x \,  B \Box \lambda \neq 0.
\end{equation}
By functional differentiation with respect to $\lambda(y)$ one obtains the
local version of (\ref{def-2-21})
\begin{equation}\label{def-2-22}
W(x) \hat{\Gamma}^{(0)}
= - \partial_\mu \frac{\delta \hat{\Gamma}^{(0)}}{\delta \tilde{A}_\mu(x)}
= \Box B(x) \neq 0.
\end{equation}
Due to the fact that this breaking is linear in the quantum field B, there do
not arise any problems for the
discussion of the gauge symmetry at higher orders of perturbation theory \cite{Boresch}, \cite{Piguet:er}.

We would like to point out that our model is
characterized  by two symmetries  at the classical level: the gauge symmetry and
the BRST-shift symmetry. This implies the 
existence of two WI's, (\ref{def-2-18})  and (\ref{def-2-21}). These WI's have
severe consequences for the computation of the 2-point vertex functional.

From eq.~(\ref{def-2-22}) follows immediately the well-known transversality
condition
\begin{equation} \label{def-2-23} 
\partial_x^\mu \frac{\delta^2\hat{\Gamma}^{(0)}}
                {\delta \tilde{A}^\mu(x)\delta\tilde{A}^\rho(y)} \Bigg |_0 = 0,
\end{equation}
where the subscript indicates vanishing classical fields.
However, the WI (\ref{def-2-18}) furnishes a further possibility to calculate
\begin{equation} \label{def-2-24}
\Pi_{\mu\rho} = \frac{\delta^2\hat{\Gamma}^{(0)}}
                {\delta \tilde{A}^\mu(x)\delta\tilde{A}^\rho(y)} \Bigg |_0.
\end{equation}
The result obtained by direct calculation (functional variation) must be consistent with the result emerging from
the  WI (\ref{def-2-18}).

However, one has to stress that all considerations
done in this section are purely classical---i.~e.\ in the tree approximation.

Therefore, one has to study the consequences of (\ref{def-2-18}).
Functional differentiation with respect to $\tilde{A}^\rho(z)$ and $c_\mu(y)$ of
$\mathcal{S}(\hat{\Gamma} ^{(0)})$ gives:
\begin{multline} \label{def-2-25}
  \frac{\delta^2}{\delta c^\mu(y) \delta \tilde{A}^\rho(z)}
  \mathcal{S}(\hat{\Gamma} ^{(0)}) \Bigg |_0 =
  \\ 
       \Big (\Box g_{\rho\sigma}
                -\partial_{\rho }\partial_{\sigma }\Big)(z)
                \frac{\delta^2\hat{\Gamma}^{(0)}}{\delta c^\mu(y) \delta
                \bar{c}_{\sigma }(z)} \Bigg |_0
        + \frac{\delta^2\hat{\Gamma}^{(0)}}
        {\delta \tilde{A}^\rho(z)\delta\tilde{A}^\mu(y)}\Bigg |_0 
  \\ 
              +\int d^4 x \, 
        \frac{\delta \mathbbm{A}^{(2)\lambda}(x)}{\delta \tilde{A}^\rho(z)} 
                \Big (\Box g_{\lambda\sigma }-\partial_\lambda\partial_{\sigma
                }\Big)(x)
                \frac{\delta^2\hat{\Gamma}^{(0)}}{\delta c^\mu(x) \delta
                \bar{c}_{\sigma }(z)} \Bigg |_0 
= 0.
\end{multline}
From (\ref{def-2-17}) one gets additionally
\begin{equation}\label{def-2-26}
                \frac{\delta^2\hat{\Gamma}^{(0)}}
                {\delta c^\mu(y) \delta\bar{c}_{\sigma }(z)} \Bigg |_0
                =
                \delta^\sigma_\mu \delta(y-z)
                +\frac{\delta  \mathbbm{A}^{(2)} _{\sigma}(z)}
                {\delta \tilde{A}^{\mu}(y)}.
\end{equation}
At order $\theta^2$ one has therefore
\begin{multline} \label{def-2-27}
         \frac{\delta^2\hat{\Gamma}^{(0)}}
                {\delta \tilde{A}^\rho(z)\delta\tilde{A}^\mu(y)} \Bigg |_0 =
\\ 
        -\Big (\Box g_{\rho\mu} - \partial_{\rho }\partial_\mu\Big)(z)
                \delta(y-z)
        -\Big (\Box g_{\rho\sigma} - \partial_{\rho }\partial_\sigma\Big)(z)
                \frac{\delta \mathbbm{A}^{(2)\sigma}(z)}{\delta \tilde{A}^\mu(y)}
\\ 
        -\Big (\Box g_{\lambda\mu}-\partial_{\lambda }\partial_\mu\Big)(y)
                \frac{\delta \mathbbm{A}^{(2)\lambda}(y)}{\delta \tilde{A}^\rho(z)} .
\end{multline}
The result (\ref{def-2-27}) is fully transversal---a consequence of the gauge
symmetry. Since $\mathbbm{A}^{(2)\mu}$ is of order $\theta^2$, (\ref{def-2-27})
yields  the well known result for the 2-point free
vertex functional for $\theta^{\rho\sigma} = 0$
\begin{equation} \label{def-2-28}
\frac{\delta^2\hat{\Gamma}^{(0)}}
        {\delta \tilde{A}^\rho(z)\delta\tilde{A}^\mu(y)} \Bigg |_0 =
        -\Big (\Box g_{\rho\mu} - \partial_{\rho }\partial_\mu\Big)(z)
        \delta(y-z).
\end{equation}
In the tree approximation there only exists a linear dependence on
$\theta^{\rho\sigma}$---therefore   $\mathbbm{A}^{(2)\mu}$ is not needed in the
action,
\begin{align}\label{def-2-29}
\Gamma _\mathrm{inv} &=
\int d^4 x \, \Big (-\frac{1}{4}F_{\mu \nu }F^{\mu \nu
}-\frac{1}{2}\theta ^{\rho \sigma }\Big (F_{\mu \rho }F_{\nu \sigma }F^{\mu \nu
}-\frac{1}{4}F_{\rho \sigma }F_{\mu \nu }F^{\mu \nu }\Big )\Big )
\nonumber \\ 
&\phantom{=}+\int d^4 x \, B\partial ^{\mu }A_{\mu },
\end{align}
which is needed to calculate eq.~(\ref{def-2-28}).

The additional terms proportional to 
$\frac{\delta  \mathbbm{A}^{(2)} _{\mu}(x)}{\delta \tilde{A}_{\sigma }(y)}$ 
in (\ref{def-2-27}) become useful if one considers one-loop corrections.

The result (\ref{def-2-27}) is easily reproduced by direct twofold functional
derivation of the action (\ref{def-2-17}) with respect to the gauge field
$\tilde{A}_\mu(x)$.

\section{$\theta$-deformed Maxwell theory: one-loop corrections}\label{3}

If one considers only the photon sector of the noncommutative Maxwell theory the
relevant action\footnote{
Here, the vectorial ghost, the antighost, the $B$ field and the external
sources are assumed to be zero.
} is given by:
\begin{align}\label{def-3-1}
\Gamma ^{(0)} &= \Gamma ^{(1)}
 + \int d^4 x \,  \partial_\mu \tilde{F}^{\mu \nu}\mathbbm{A}^{(2)}_{\nu } 
 \nonumber \\
&= \int d^4 x \, \Big (-\frac{1}{4}\Big (\tilde{F}_{\mu \nu }\tilde{F}^{\mu \nu }
-4\partial ^{\mu }\tilde{F}_{\mu \nu } \mathbbm{A} ^{(2)\nu }\Big ) \nonumber \\
&\phantom{=} -\frac{1}{2}\theta ^{\rho \sigma }\Big (\tilde{F}_{\mu \rho }\tilde{F}_{\nu
\sigma }\tilde{F}^{\mu \nu }-\frac{1}{4}\tilde{F}_{\rho \sigma }
\tilde{F}_{\mu\nu }\tilde{F}^{\mu \nu }\Big )\Big ),
\end{align}
where $\Gamma^{(1)}$ denotes terms of order 0 and 1 in $\theta$.
Eq.~(\ref{def-3-1}) follows from (\ref{def-2-17}).
In order to compensate the one-loop selfenergy corrections one needs the
explicit form of $\mathbbm{A} ^{(2)}_\nu $ \cite{Bichl:2001cq} :
\begin{align}\label{def-3-2}
\mathbbm{A}^{(2)}_\mu
&=\phantom{+}
  \kappa^{(2)}_1
        g^{\alpha\gamma}g^{\beta\delta}g^{\lambda\rho}g^{\sigma\tau}
        \theta_{\alpha\beta}\theta_{\gamma\delta}
        \partial_\lambda\partial_\rho\partial_\sigma
        \tilde{F}_{\tau\mu}
 \nonumber \\
&\phantom{=}+ \kappa^{(2)}_2
        g^{\alpha\gamma}g^{\beta\lambda}g^{\delta\rho}g^{\sigma\tau}
        \theta_{\alpha\beta}\theta_{\gamma\delta}
        \partial_\lambda\partial_\rho\partial_\sigma
        \tilde{F}_{\tau\mu}
 \nonumber \\
&\phantom{=}+ \kappa^{(2)}_3 
        g^{\beta\sigma}g^{\gamma\tau}g^{\alpha\lambda}g^{\delta\rho}
        \theta_{\mu\beta}\theta_{\gamma\delta}
        \partial_\alpha\partial_\lambda\partial_\rho
        \tilde{F}_{\sigma\tau}
 \nonumber \\
&\phantom{=}+ \kappa^{(2)}_4 
        g^{\gamma\tau}g^{\beta\delta}g^{\alpha\lambda}g^{\rho\sigma}
        \theta_{\mu\beta}\theta_{\gamma\delta}
        \partial_\alpha\partial_\lambda\partial_\rho
        \tilde{F}_{\sigma\tau},
\end{align}
which is  gauge invariant.

Inserting (\ref{def-3-2}) into (\ref{def-3-1}) one gets for the terms quadratic in
$\theta^{\mu\nu}$
\begin{multline}\label{def-3-3}
\Gamma ^{(0)} = \Gamma ^{(1)} +\int d^{4} x \, \tilde{A}_\mu\Bigg(
         \Big(g^{\mu\nu}\Box-\partial^\mu\partial^\nu \Big)
        \Big( \kappa^{(2)}_1 \theta^2\Box^2 
        + \kappa^{(2)}_2 \Tilde{\Tilde{\overset{}{\Box}}}\Box \Big) \\
        + \kappa^{(2)}_3 \tilde{\partial}^\mu \tilde{\partial}^\nu \Box^2
        + \kappa^{(2)}_4 \Big(\theta^{\mu\alpha}\theta^\nu_\alpha \Box^3
                + \Big( \Tilde{\Tilde{\partial}}^\mu\partial^\nu 
                + \Tilde{\Tilde{\partial}}^\nu\partial^\mu \Big) \Box^2
                + \partial^\mu\partial^\nu \Tilde{\Tilde{\overset{}{\Box}}} \Box
                \Big)
        \Bigg)\tilde{A}_\nu,
\end{multline}
where $\Box=\partial^\alpha\partial_\alpha$,
$\tilde{\partial}^\alpha=\theta^{\alpha\beta}\partial_\beta$,
$\Tilde{\Tilde{\partial}}^\alpha=\theta^{\alpha\beta}\tilde{\partial}_\beta$,
$\Tilde{\Tilde{\overset{}{\Box}}}=\tilde{\partial}^\alpha\tilde{\partial}_\alpha$ and
$\theta^2=\theta^{\alpha\beta}\theta_{\alpha\beta}$.

With the help of (\ref{def-3-2}) one verifies by direct calculation that
(\ref{def-3-3}) and (\ref{def-2-27}) are consistent.

At the one-loop level this means that the shift symmetry controls the radiative
corrections of the perturbative calculation.

In order to cancel the one-loop divergences one performs the following
renormalization of $\kappa^{(2)}_1$, $\kappa^{(2)}_2$, $\kappa^{(2)}_3$ and
$\kappa^{(2)}_4$ \cite{Bichl:2001cq}:
\begin{alignat}{2}\label{def-3-4}
\kappa^{(2)}_1 \; & \rightarrow  \; \kappa^{(2)}_1
-\frac{1}{16}\frac{\hbar}{(4\pi)^2 \epsilon} \, ,
\qquad &
\kappa^{(2)}_2 \; & \rightarrow  \; \kappa^{(2)}_2
+\frac{1}{20}\frac{\hbar}{(4\pi)^2 \epsilon} \, ,
\\ \nonumber
\kappa^{(2)}_3 \; & \rightarrow  \; \kappa^{(2)}_3
+\frac{1}{60}\frac{\hbar}{(4\pi)^2 \epsilon} \, , 
\qquad &
\kappa^{(2)}_4 \; & \rightarrow  \; \kappa^{(2)}_4
+\frac{1}{8}\frac{\hbar}{(4\pi)^2 \epsilon}\, .
\end{alignat}
However, one has to stress that (\ref{def-3-4}) represent unphysical
renormalizations because the $\kappa^{(2)}_i$ parametrize the field redefinition
(\ref{def-2}) and (\ref{def-2-10}).

\section{Conclusion}\label{4}

In this paper we have demonstrated the usefulness of the BRST-shift symmetry in
connection with the renormalization program of the vacuum polarization of the
$\theta$-deformed Maxwell theory at the one-loop level. Gauge symmetry and
BRST-shift symmetry can be implemented consistently. Unfortunately the non-Abelian extension is plagued by several difficulties.


\begin{thebibliography}{99}

\bibitem{Moyal:sk}
J.~E.~Moyal,
``Quantum Mechanics As A Statistical Theory,''
Proc.\ Cambridge Phil.\ Soc.\  {\bf 45} (1949) 99.

\bibitem{Hayakawa:1999zf}
M.~Hayakawa,
``Perturbative analysis on infrared and ultraviolet aspects of  noncommutative QED on R**4,''
arXiv:hep-th/9912167.

\bibitem{Hayakawa:1999yt}
M.~Hayakawa,
``Perturbative analysis on infrared aspects of noncommutative QED on  R**4,''
Phys.\ Lett.\ B {\bf 478} (2000) 394
[arXiv:hep-th/9912094].

\bibitem{Matusis:2000jf}
A.~Matusis, L.~Susskind and N.~Toumbas,
``The IR/UV connection in the non-commutative gauge theories,''
JHEP {\bf 0012} (2000) 002
[arXiv:hep-th/0002075].

\bibitem{x}
Work in preparation

\bibitem{Seiberg:1999vs}
N.~Seiberg and E.~Witten,
``String theory and noncommutative geometry,''
JHEP {\bf 9909} (1999) 032
[arXiv:hep-th/9908142].


\bibitem{Madore:2000en}
J.~Madore, S.~Schraml, P.~Schupp and J.~Wess,
``Gauge theory on noncommutative spaces,''
Eur.\ Phys.\ J.\ C {\bf 16} (2000) 161
[arXiv:hep-th/0001203].


\bibitem{Jurco:2000ja}
B.~Jurco, S.~Schraml, P.~Schupp and J.~Wess,
``Enveloping algebra valued gauge transformations for non-Abelian gauge  groups on non-commutative spaces,''
Eur.\ Phys.\ J.\ C {\bf 17} (2000) 521
[arXiv:hep-th/0006246].

\bibitem{Jurco:2001rq}
B.~Jurco, L.~Moller, S.~Schraml, P.~Schupp and J.~Wess,
``Construction of non-Abelian gauge theories on noncommutative spaces,''
Eur.\ Phys.\ J.\ C {\bf 21} (2001) 383
[arXiv:hep-th/0104153].

\bibitem{Bichl:2001nf}
A.~A.~Bichl, J.~M.~Grimstrup, L.~Popp, M.~Schweda and R.~Wulkenhaar,
``Perturbative analysis of the Seiberg-Witten map,''
arXiv:hep-th/0102044.

\bibitem{Alfaro:1992cs}
J.~Alfaro and P.~H.~Damgaard,
``BRST symmetry of field redefinitions,''
Annals Phys.\  {\bf 220} (1992) 188.


\bibitem{Kruglov:2001dm}
S.~I.~Kruglov,
``Maxwell's theory on non-commutative spaces and quaternions,''
arXiv:hep-th/0110059.


\bibitem{Jackiw:2001dj}
R.~Jackiw,
``Physical instances of noncommuting coordinates,''
arXiv:hep-th/0110057.


\bibitem{Guralnik:2001ax}
Z.~Guralnik, R.~Jackiw, S.~Y.~Pi and A.~P.~Polychronakos,
``Testing non-commutative QED, constructing non-commutative MHD,''
Phys.\ Lett.\ B {\bf 517} (2001) 450
[arXiv:hep-th/0106044].

\bibitem{Cai:2001az}
R.~G.~Cai,
``Superluminal noncommutative photons,''
Phys.\ Lett.\ B {\bf 517} (2001) 457
[arXiv:hep-th/0106047].


\bibitem{Bichl:2001gu}
A.~A.~Bichl, J.~M.~Grimstrup, L.~Popp, M.~Schweda and R.~Wulkenhaar,
``Deformed QED via Seiberg-Witten map,''
arXiv:hep-th/0102103.


\bibitem{Bichl:2001cq}
A.~Bichl, J.~Grimstrup, H.~Grosse, L.~Popp, M.~Schweda and R.~Wulkenhaar,
``Renormalization of the noncommutative photon selfenergy to all orders via Seiberg-Witten map,''
JHEP {\bf 0106} (2001) 013
[arXiv:hep-th/0104097].


\bibitem{Wulkenhaar:2001sq}
R.~Wulkenhaar,
``Non-renormalizability of theta-expanded noncommutative QED,''
arXiv:hep-th/0112248.

\bibitem{Bastianelli:1990ey}
F.~Bastianelli,
``BRST Symmetry From A Change Of Variables And The Gauged WZNW Models,''
Nucl.\ Phys.\ B {\bf 361} (1991) 555.

\bibitem{Boresch}
A.~Boresch, S.~Emery, O.~Moritsch, M.~Schweda, T.~Sommer, H.~Zerrouki,
``Applications of Noncovariant Gauges in the Algebraic Renormalization Procedure,''
{\it Singapore, Singapore: World Scientific (1998) 150 p}.


\bibitem{Piguet:er}
O.~Piguet and S.~P.~Sorella,
``Algebraic Renormalization: Perturbative Renormalization, Symmetries And Anomalies,''
Lect.\ Notes Phys.\  {\bf M28} (1995) 1.




\end{thebibliography}
\end{document}